\documentclass{article}
\usepackage{spconf,amsmath,graphicx}

\usepackage{booktabs}
\usepackage{float}
\usepackage[T1]{fontenc}
\usepackage{amsfonts}
\usepackage{bm}
\usepackage{algorithm}
\usepackage[noend]{algpseudocode}
\usepackage{amsmath}
\usepackage{amsfonts}
\usepackage{cite}
\usepackage{multirow}

\title{MULTIPLE CONFIDENCE GATES FOR JOINT TRAINING OF SE AND ASR}

\name{Tianrui Wang$^{\star \dagger}$, Weibin Zhu$^{\star}$, Yingying Gao$^{\dagger}$, Junlan Feng$^{\dagger}$, Shilei Zhang$^{\dagger}$}
\address{$^{\star}$ Institute of Information Science, Beijing Jiaotong University, Beijing, China \\
	$^{\dagger}$ China Mobile Research Institute, Beijing, China}
\begin{document}
	%\ninept
	%
	\maketitle

	\begin{abstract}
		Joint training of speech enhancement model (SE) and speech recognition model (ASR) is a common solution for robust ASR in noisy environments. SE focuses on improving the auditory quality of speech, but the enhanced feature distribution is changed, which is uncertain and detrimental to the ASR. To tackle this challenge, an approach with multiple confidence gates for jointly training of SE and ASR is proposed. A speech confidence gates prediction module is designed to replace the former SE module in joint training. The noisy speech is filtered by gates to obtain features that are easier to be fitting by the ASR network. The experimental results show that the proposed method has better performance than the traditional robust speech recognition system on test sets of clean speech, synthesized noisy speech, and real noisy speech.
	\end{abstract}
	\section{Introduction}
	\label{sec:intro}
	The performance of ASR in a noisy environment will greatly deteriorate, and it has been a persistent and challenging task to improve the adaptability of ASR in a noisy environment \cite{intro1}. 
	
	From the perspective of features, some researchers have proposed feature construction strategies that are more robust to noise, in order to reduce the difference in features caused by noise and thereby improve the robustness of ASR \cite{introfeature1,introfeature2,introfeature3}. But these methods are difficult to work with a low signal-to-noise ratio (SNR). From the perspective of model design, a direct approach is to set a SE in front of the ASR \cite{se2asr1}. Although the enhanced speech is greatly improved currently in human hearing \cite{Nsnet,crn,dccrn,apcsnr}, the enhanced feature distribution is changed also, which may be helpful for the hearing but not always necessarily beneficial for ASR. In order to make the ASR adaptable to the enhanced feature, the ASR can be retrained with the data processed by the enhanced model \cite{se2asr2}. However, the performance of this method is severely affected by the SE effect. Especially in the case of low SNR, the SE may corrupt the speech structure and bring somehow additional noise. In order to reduce the influence of SE and achieve better recognition accuracy, researchers proposed the joint training strategy \cite{introframework}. Let ASR constrain SE and train the model with the recognition accuracy of ASR as the main goal. However, the models are difficult to converge during joint training, and the final recognition performance improvement is limited due to the incompatibility between two goals, for SE it's speech quality, while for ASR it's recognition accuracy.
	
	% During joint training, the tuning of the signal distribution by the enhancement model is always changing, so it’s difficult to converge for ASR, which aims to fit the signal distribution. 
	During joint training, the ASR network needs to continuously fit the feature distribution, but the feature distribution keeps changing under the action of SE, which makes it difficult for the ASR network to converge. We believe that the ASR network has strong fitting and noise-carrying capabilities. We need neither to do too much processing on the feature distribution nor to predict the value of clean speech at each feature point. We only need to filter out the feature points that don't contain speech. This method can avoid the problem that ASR cannot converge during joint training, and also give full play to the noise-carrying ability of ASR itself. So, the multiple confidence gates for joint training of SE and ASR is proposed. A speech confidence gates prediction module is designed to replace the former speech enhancement module in joint training. Each gate is a confidence spectrum with the same size as the feature spectrum (the probability that each feature point contains speech). The noisy speech is filtered by each confidence gate respectively. Then the gated results are combined to obtain the input feature for ASR. The experimental results show that the proposed method can effectively improve the recognition accuracy of the joint training model, and has better performance than the traditional robust ASR on test sets.

	\begin{figure}[htb]
		\centering
		\vspace{-0.7cm}
		\includegraphics[width=8.0cm]{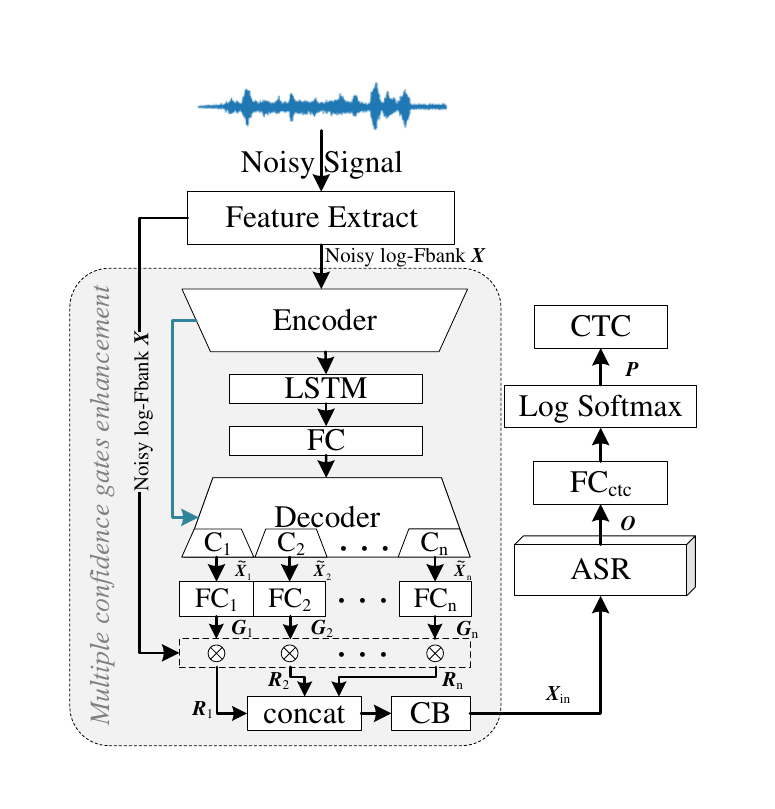}
		\vspace{-0.7cm}
		\caption{Architecture of the proposed method.}
		\label{allmodel}
		\vspace{-0.4cm}
	\end{figure}

	\section{Proposed method}
	The overall diagram of the proposed system is shown in Fig.\ref{allmodel}. It is mainly comprised of two parts, namely the multiple confidence gates enhancement module (MCG) and ASR. Firstly, We convert the waveform to logarithmic Fbank (log-Fbank)  $\bm{X} \in \mathbb{R}^{T\times Q}$ as the input to the model by the short-time fast Fourier transform (STFT) \cite{stft} and auditory spectrum mapping \cite{fbank}, where $T$ and $Q$ denote the number of frames and the dimension of the Bark spectrum respectively. Secondly, The MCG predicts multiple confidence gates based on the noisy log-Fbank, and the noisy features are filtered by each gate respectively. Then the filtered results are combined into inputs of ASR by a convolution block (CB). Different confidence gates correspond to the selections of speech feature points with different energy thresholds, and each process will be described later. 
	
	\subsection{Multiple confidence gates enhancement module}
	\label{sec:SPM}
	The multiple confidence gates enhancement module is used to predict the confidence gates to filter the noisy spectrum. The main body of MCG is the convolutional recurrent network (CRN) \cite{crn}, it is an encoder-decoder architecture for speech enhancement. Both the encoder and decoder are comprised of convolution blocks (CB). And each CB is comprised of 2D convolution (2d-Conv), batch normalization (BN) \cite{batchnorm}, and parametric rectified linear unit (PReLU) \cite{prelu}. Between the encoder and decoder, long short-term memory (LSTM) layer \cite{lstm} is inserted to model the temporal dependencies. Additionally, skip connections are utilized to concatenate the output of each encoder layer to the input of the corresponding decoder layer (blue line in Fig.\ref{allmodel}). 

	Since we want to determine the confidence that each feature point contains speech, we change the output channel number of the last decoder to $(\text{C}_1+\text{C}_2+\cdots +\text{C}_n)$, where $\text{C}_1, \text{C}_2, \cdots, \text{C}_n$ are the number of channels of the input features for corresponding fully connected layers $(\text{FC}_1,\text{FC}_2,\cdots,\\ \text{FC}_n)$. Each FC can output a confidence gate $\bm{G} \in \mathbb{R}^{T \times Q}$ for every feature point,
	\begin{equation}
		\label{confidence}
		\setlength{\abovedisplayskip}{3pt}
		\setlength{\belowdisplayskip}{3pt}
		\bm{G}_n = \text{sigmoid}(\tilde{\bm{X}}_n \cdot \bm{W}_n + b_n)
	\end{equation}
	where $\tilde{\bm{X}}_n \in \mathbb{R}^{T\times Q \times \text{C}_n}$ represents the $\text{C}_n$ channels of the decoder output. $\bm{W}_n \in \mathbb{R}^{\text{C}_n \times 1}$ and $b_n$ are the parameters of $\text{FC}_n$. $\text{sigmoid}$ \cite{sigmoid} is used to convert the result into confidence (the probability that a feature point contains speech).
	
\begin{figure}[htb]
		\centering
		\vspace{-0.4cm}
		\includegraphics[width=8cm]{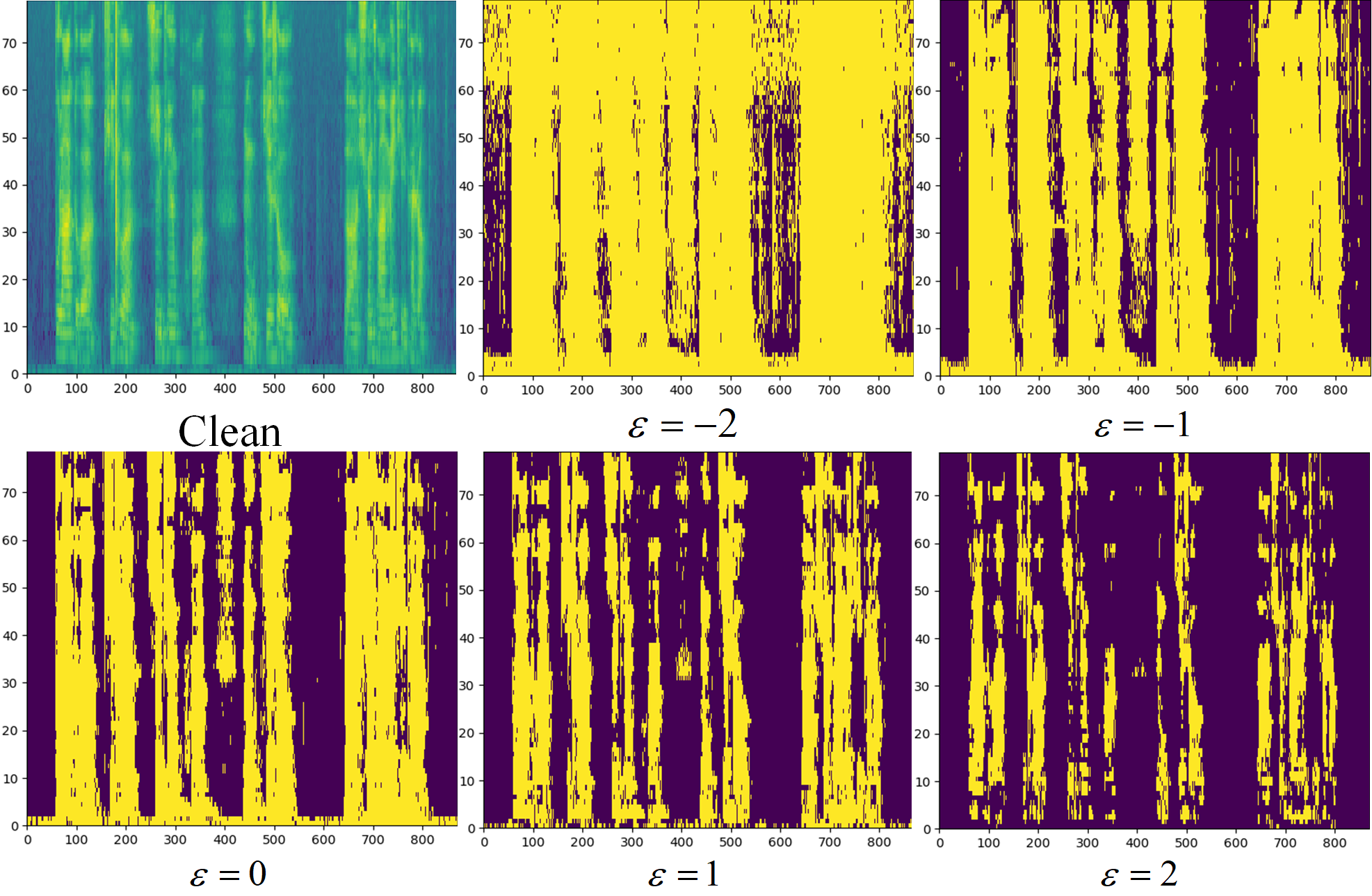}
		\vspace{-1em}
		\caption{$\dot{\bm{G}}$ obtained by different $\varepsilon$. The greater $\varepsilon$, the greater the speech energy of the corresponding choosed points.}
		\label{E}
	\end{figure}

	The label of confidence gate $\bm{G}$ is designed based on energy statistics. Different thresholds are used to control different filtering degrees. The greater the energy threshold, the fewer the number of speech points in the label, and the greater the speech energy of the corresponding feature points, as shown in Fig.\ref{E}. We first compute the mean $\bm{\mu} \in \mathbb{R}^{Q \times 1}$ of the log-Fbank of the clean speech set $\bm{\mathbb{X}}=[\dot{\bm{X}}_1,\cdots,\dot{\bm{X}}_d] \in \mathbb{R}^{D\times T \times Q}$ and standard deviation $\bm{\sigma} \in \mathbb{R}^{Q \times 1}$ of log-Fbank means, 
	\begin{equation}
		\setlength{\abovedisplayskip}{3pt}
		\setlength{\belowdisplayskip}{3pt}
		\bm{\mu} = \sum_{i=1}^{D}\left[\left(\sum_{t=1}^{T}\bm{\mathbb{X}}_{i,t}\right)/T\right]/D
	\end{equation}
	\begin{equation}
		\setlength{\abovedisplayskip}{3pt}
		\setlength{\belowdisplayskip}{3pt}
		\bm{\sigma} = \sqrt{\sum_{i=1}^{D}\left[\left(\sum_{t=1}^{T}\bm{\mathbb{X}}_{i,t}\right)/T-\bm{\mu}\right]^2/D}
	\end{equation}
	where $\dot{\bm{X}}$ represents log-Fbank of the clean speech. $D$ represents the clip number of audio in clean speech set $\bm{\mathbb{X}}$. Then the thresholds $\bm{\kappa}=(\bm{\mu}+\varepsilon \cdot \bm{\sigma})\in \mathbb{R}^{Q \times 1}$ for bins are controlled according to different offset values $\varepsilon$, and the confidence gate label $\dot{\bm{G}}$ is 1 if the feature point value is larger than $\bm{\kappa}$, 0 otherwise, as shown in Fig.\ref{E}.
	\begin{equation}
		\setlength{\abovedisplayskip}{3pt}
		\setlength{\belowdisplayskip}{3pt}
		\label{confidencelabel}
		\dot{\bm{G}}_{t,q} = 
		\left\{
		\begin{array}{rcl}
			& 1 &, \dot{\bm{X}}_{t,q} \ge \bm{\kappa}  \\
			& 0 &, \dot{\bm{X}}_{t,q} < \bm{\kappa}
		\end{array}
		\right.
	\end{equation}

	After (\ref{confidence}), we can get $n$ confidence gates $\left[\bm{G}_1,\cdots,\bm{G}_n \right]$ corresponding to different $\varepsilon$. Then noisy log-Fbank are filtered by the confidence gates, resulting in different filtered results $\left[\bm{R}_1,\cdots,\bm{R}_n\right]$,
	\begin{equation}
		\setlength{\abovedisplayskip}{3pt}
		\setlength{\belowdisplayskip}{3pt}
		\label{selectedregion}
		\bm{R}_n = \bm{G}_n \otimes \bm{X}
	\end{equation}
	where $\otimes$ is the element-wise multiplication. And these filtered results are then concatenated in channel dimension and input to a CB to obtain the input $\bm{X}_\text{in}$ of ASR.

	\subsection{Automatic speech recognition}
	The fitting capability of existing ASR network is very powerful. Since the Conformer+Transformer model is too large for joint training, we only use Conformer encoder \cite{conformer} as ASR module. This model first processes the enhanced input $\bm{X}_\text{in} \in \mathbb{R}^{T \times Q}$ with a convolution subsampling layer (CSL),
	\begin{equation}
		\setlength{\abovedisplayskip}{3pt}
		\setlength{\belowdisplayskip}{3pt}
		\bm{X}_\text{down} =  \text{CSL}\left(\bm{X}_\text{in}\right)
	\end{equation}
	where CSL is comprised of subsampling convolution, linear layer, and dropout. Then the downsampled feature is processed by several conformer blocks. Each conformer block is comprised of four modules stacked together, i.e, a half-step residual feed-forward module (FFN), a self-attention module (MHSA), a convolution module ($\text{Conv}_c$), and a second FFN is followed by a layer norm \cite{layernorm}  in the end,
	\begin{equation}
		\setlength{\abovedisplayskip}{3pt}
		\setlength{\belowdisplayskip}{3pt}
		\bm{X}_i^{'} = \bm{X}_i + \frac{1}{2} \text{FFN}(\bm{X}_i) 
	\end{equation}
	\begin{equation}
		\setlength{\abovedisplayskip}{3pt}
		\setlength{\belowdisplayskip}{3pt}
		\bm{X}_i^{''} = \bm{X}_i^{'} + \text{MHSA}(\bm{X}_i^{'})
	\end{equation}
	\begin{equation}
		\setlength{\abovedisplayskip}{3pt}
		\setlength{\belowdisplayskip}{3pt}
		\bm{X}_i^{'''}=\bm{X}_i^{''}+\text{Conv}_c(\bm{X}_i^{''})
	\end{equation}
	\begin{equation}
		\setlength{\abovedisplayskip}{3pt}
		\setlength{\belowdisplayskip}{3pt}
		\bm{O}_i = \text{Layernorm}(\bm{X}_i^{'''}+\frac{1}{2}\text{FFN}(\bm{X}_i^{'''}))
		\label{O}
	\end{equation}
	where $\bm{X}_i$ represents the input of $i$-th Conformer block ($\bm{X}_0=\bm{X}_\text{down}$), and the $\bm{O}_i$ is the output of the block. The final output of Conformer is denoted as $\bm{O} \in \mathbb{R}^{T\times H}$, where $H$ represents the hidden dimension of the last layer.

	\subsection{Loss function}
	The loss function of our joint training framework is comprised of four components. 
	\begin{equation}
		\setlength{\abovedisplayskip}{3pt}
		\setlength{\belowdisplayskip}{3pt}
		L =  L_G + L_R + L_\text{O} + L_\text{CTC}
	\end{equation}
	
	In the MCG module, we measured the predicted confidence gate as,
	\begin{equation}
		\setlength{\abovedisplayskip}{3pt}
		\setlength{\belowdisplayskip}{3pt}
		L_G = \sum_{i=1}^{n}||\bm{G}_n-\dot{\bm{G}}_n||_1
	\end{equation}
	where $||\cdot||_1$ represents the 1 norm operation. In addition, to strengthen the filtering ability of the module on noise, we also compute filtered results from the clean speech through the same processing, which are denoted as $\left[\dot{\bm{R}}_1,\cdots,\dot{\bm{R}}_n\right]$, and calculate the difference between them and the filtered results of noisy speech as,
	\begin{equation}
		\setlength{\abovedisplayskip}{3pt}
		\setlength{\belowdisplayskip}{3pt}
		L_R = \sum_{i=1}^{n}||\bm{R}_n-\dot{\bm{R}}_n||_1
	\end{equation}
	
	After the ASR, In order to reduce the noise-related changes brought by SE to ASR, we also measured the difference between the $\dot{\bm{O}}$ computed from clean speech and that from noisy speech as, 
	\begin{equation}
		\setlength{\abovedisplayskip}{3pt}
		\setlength{\belowdisplayskip}{3pt}
		L_\text{O} = ||\bm{O}-\dot{\bm{O}}||_1
	\end{equation}

	Note that the gradients of the processing of all clean speech are discarded. And $L_\text{CTC}$ is the connectionist temporal classification \cite{ctc} for ASR.
	
	\section{EXPERIMENTS}
	\subsection{Dataset}
	We use AISHELL1 \cite{aishell1} as the clean speech dataset, which includes a total of 178 hours of speech from 400 speakers, and it is divided into training, development, and testing sets. The noise data used in training, development, and test set are from DNS challenge (INTERSPEECH 2020) \cite{dns2020}, ESC50 \cite{esc50}, and  Musan \cite{musan}, respectively. During the mixing process, a random noise cut is extracted and then mixed with the clean speech under an SNR selected from -5dB to 20dB. In addition, we randomly selected 2600 speech clips from the noisy speech recorded in real scenarios as an additional test set, and call it APY \footnote{https://www.datatang.com/dataset/info/speech/191}.
%	\begin{table*}[htp]
%		\setlength\tabcolsep{2.7pt}
%		\centering
%		\caption{Recognition results of the models on test set}
%		\begin{tabular}{l|ccccc|ccccc|ccccc}
%			\toprule
%			\multirow{2}*{Model} & \multicolumn{5}{|c|}{Clean}& \multicolumn{5}{|c|}{Noisy}&\multicolumn{5}{|c}{APY} \\
%			\cline{2-16}
%			& C & S  & D & I & WER & C & S  & D & I & WER & C & S  & D & I & WER  \\
%			\midrule
%			Conformer   & 12.998 & 1.580 & 0.050 & 0.039 & 11.580 & 11.659 & 2.814 & 0.154 & 0.075 & 20.984 & 4.500 & 2.730  & 0.206 & 0.181 & 44.969 \\	
%			\ \ +Nsnet   & 13.145 & 1.440 & 0.042 & \textbf{0.026} & 10.427 & 11.607 & 2.859 & 0.161 & 0.083 & 21.302 & 4.018 & 3.188 & \textbf{0.166} & 0.213 & 51.465   \\
%			\ \ +DCRN  & 13.154 & 1.428  & 0.045 & 0.027 & 10.333 & 11.675 & 2.750 & 0.202 & 0.069 & 20.486 & 4.220 & 2.977 & 0.174 & 0.220 & 48.820   \\
%			\ \ +DCCRN & 13.151 & 1.429  & 0.047 & 0.028 & 10.422 & 11.782 & 2.652 & 0.194 & 0.063 & 19.904 & 4.314 & 2.868  & 0.190 & 0.240 & 47.705  \\
%			\ \ +DCRN(ST)  & 13.141 & 1.440  & 0.046 & 0.029 & 10.449 & 11.656 & 2.739 & 0.232 & 0.059 & 20.844 & 4.068 & 3.154 & 0.214 & 0.180 & 50.529   \\
%			\ \ +\textbf{MCG}  & \textbf{13.313} & \textbf{1.271}  & \textbf{0.041} & 0.030 & \textbf{9.343} & \textbf{12.231} & \textbf{2.256} & \textbf{0.140} & \textbf{0.056} & \textbf{16.882} & \textbf{4.807} & \textbf{2.391} & 0.174 & \textbf{0.125} & \textbf{39.223} \\
%			
%			\bottomrule
%		\end{tabular}
%		\label{tab:1}
%		\vspace{-0.4cm}
%	\end{table*}

	\begin{table*}[htp]
		\centering
		\caption{Recognition results of the models on test set}
		\begin{tabular}{l|cccc|cccc|cccc}
			\toprule
			\multirow{2}*{Model} & \multicolumn{4}{|c|}{Clean}& \multicolumn{4}{|c|}{Noisy}&\multicolumn{4}{|c}{APY} \\
			\cline{2-13}
			& S  & D & I & WER  & S  & D & I & WER  & S  & D & I & WER  \\
			\midrule
			Conformer   & 1.580 & 0.050 & 0.039 & 11.580  & 2.814 & 0.154 & 0.075 & 20.984  & 2.730  & 0.206 & 0.181 & 44.969 \\	
			\ \ +Nsnet    & 1.440 & 0.042 & \textbf{0.026} & 10.427 & 2.859 & 0.161 & 0.083 & 21.302 & 3.188 & \textbf{0.166} & 0.213 & 51.465   \\
			\ \ +DCRN   & 1.428  & 0.045 & 0.027 & 10.333 & 2.750 & 0.202 & 0.069 & 20.486 & 2.977 & 0.174 & 0.220 & 48.820   \\
			\ \ +DCCRN  & 1.429  & 0.047 & 0.028 & 10.422  & 2.652 & 0.194 & 0.063 & 19.904 & 2.868  & 0.190 & 0.240 & 47.705  \\
			\ \ +DCRN(ST)  & 1.440  & 0.046 & 0.029 & 10.449 & 2.739 & 0.232 & 0.059 & 20.844 & 3.154 & 0.214 & 0.180 & 50.529   \\
			\ \ +\textbf{MCG}  & \textbf{1.271}  & \textbf{0.041} & 0.030 & \textbf{9.343}  & \textbf{2.256} & \textbf{0.140} & \textbf{0.056} & \textbf{16.882} & \textbf{2.391} & 0.174 & \textbf{0.125} & \textbf{39.223} \\
			\bottomrule
		\end{tabular}
		\label{tab:1}
		\vspace{-0.5cm}
	\end{table*}

	\subsection{Training setup and baseline}
	\label{trainsetup}
	We trained all models on our dataset with the same strategy. The initial learning rate is set to 0.0002, which will decay 0.5 when the validation loss plateaued for 5 epochs. The training is stopped if the validation loss plateaued for 20 epochs. And the optimizer is Adam \cite{adam}. We design a Conformer, three traditional joint training systems, and a separately training (ST) system as the baselines.
	
	\textbf{Conformer}: The dimension of log-Fbank is 80. The number of conformer blocks is 12. Linear units is 2048. The convolution kernel is 15. Attention heads is 4. Output size is 256. And we replaced the CMVN with BN to help Conformer integrate with the SE module.
	
	\textbf{Nsnet+Conformer}: Nsnet is a lightweight and effective speech enhancement model. It's mainly comprised of FC and gated recurrent units (GRUs). The Nsnet we implemented is the same structure as it in \cite{Nsnet}. 
	
	\textbf{DCRN+Conformer}: DCRN is an encoder-decoder model for speech enhancement in the complex domain \cite{crn}. The $32\text{ms}$ Hanning window with $25\%$ overlap and 512-point STFT are used. The channel number of encoder and decoder is $\left\{32,64,128,256,256,256\right\}$. Kernel size and stride are (5,2) and (2,1). One 128-units LSTM is adopted. And a 1024-units FC layer is after the LSTM.
	
	\textbf{DCRN+Conformer(ST)}: Separately training is to train a DCRN first, then the conformer is trained based on the processed data by the DCRN. The parameter configuration is the same as that of the joint training.
	
	\textbf{DCCRN+Conformer}: DCCRN is the complex operation version of DCRN with stronger speech enhancement effect \cite{dccrn}. The $25\text{ms}$ Hanning window with $25\%$ overlap and 512-point STFT are used. The channel number is $\left\{32,64,128,256,256,256\right\}$. Kernel size and stride are (5,2) and (2,1). And it uses two layers complex LSTM with 128 units for the real part and imaginary part respectively. And a dense layer with 1280 units is after the LSTM.
	
	\textbf{MCG+Conformer}: The channel numbers of MCG are $\left\{32,48,64,80,96\right\}$. Kernel size of each convolution is (3,3), strides are $\left\{(1,1),(1,1),(2,1),(2,1),(1,1),(1,1)\right\}$. One 128-units LSTM is adopted. And a 1920-units FC layer is after the LSTM. The channel number of the last decoder is $10n$ and $\text{C}_1=\cdots=\text{C}_n=10$. The best results are obtained when $n=3$ and $\varepsilon=\left[-1,1,2\right]$.
	
	\begin{table}[htp]
		\centering
		\vspace{-0.4cm}
		\caption{Test results of enhancement models in different baseline systems on test sets}
		\begin{tabular}{cccc}
			\toprule
			Model &  PESQ & STOI(\%)  & SI-SDR(dB) \\ 
			\midrule
			Noisy  & 1.643 & 0.796 &  7.218 \\	
			Nsnet   & 2.373  & 0.857   &  15.312   \\
			DCRN   & 2.508  & 0.858   &  15.232   \\
			DCCRN  & 2.570 & 0.868  & 16.004  \\
			DCRN(ST)   & \textbf{2.617}  & \textbf{0.871}   &  \textbf{16.441}   \\
			\bottomrule
			\label{results}
		\end{tabular}
		\label{tab:2}
		\vspace{-0.6cm}
	\end{table}

	\subsection{Experimental results and discussion}
	We trained the model as described in Section \ref{trainsetup}, and counted the substitution error (S), deletion error (D), insertion error (I), and word error rate (WER) of different systems on the clean, noisy, and APY test set, shown as Table \ref{tab:1}. And Table \ref{tab:2} shows the effects of the enhancement modules in different baseline systems. The enhancement performance is measured by the PESQ, STOI, and SI-SDR, which are commonly used in speech enhancement. The PESQ is a measure for magnitude difference in the range of human hearing, the STOI is more concerned with the correlation between the enhanced magnitude and the referenced one, and the SI-SDR is more concerned with the structural difference in the time-domain.
	
	It can be seen from the results of Table \ref{tab:1} and Table \ref{tab:2}, among the three joint training baselines, the best performance is achieved when DCCRN is used as the front-end, which proves that the robustness of recognition will be improved as the performance of the SE increases. However, on the APY dataset, the WER of the baselines are all higher than those of the Conformer. Because the SE is over-fitting on the synthesized data, and in the real scenarios, the noise is more complicated, which increases the uncertainty of feature changes brought by SE.
	
	The enhancement performance of separately trained DCRN is better than that of joint training, but the recognition accuracy of separately training is worse than that of joint training. Because joint training is mainly aimed at recognition, the performance of SE will be limited.
	
	Our method achieved the lowest WER on all test sets, moreover, the insertion and substitution errors of our model are minimal on the noisy test set, which demonstrates that the multiple confidence gates scheme can filter out the non-speech parts without destroying the speech structure, thereby greatly improving the recognition performance. The performance improvement on APY also indicates that the change of feature by the proposed front-end module is less affected by noise and more friendly to ASR.

	\begin{table}[htp]
		\setlength\tabcolsep{4pt}
		\centering
		\vspace{-0.4cm}
		\caption{Recognition results models trained by different confidence gates on test set}
		\begin{tabular}{ccccc}
			\toprule
			n & $\varepsilon$ & Clean WER & Noisy WER & APY WER  \\
			\midrule
			1 & $\left[0\right]$  & 9.457 & 17.134 & 39.361 \\	
			2 & $\left[-1,1\right]$  & 9.669 & 17.626 & 41.211   \\
			3  & $\left[-1,1,2\right]$ & \textbf{9.343} & \textbf{16.882} & \textbf{39.223}   \\
			4  & $\left[-2,-1,1,2\right]$ & 9.498 & 17.440 & 40.513 \\
			\bottomrule
		\end{tabular}
		\label{tab:3}
		\vspace{-0.1cm}
	\end{table}

	In addition, we have done some experiments on the design of the confidence gates. We set different thresholds to control the filtering ability of the gates. It can be seen that the best performance of the model is achieved with $n=3$ and $\varepsilon=\left[-1,1,2\right]$. As can be seen from Fig.\ref{E}, as $\varepsilon$ increases, the higher the energy of selected feature points, the stronger the filtering ability of the gate. So designing multiple gates with different filtering abilities can improve the fitting and generalization ability of model to achieve better results.

	\section{CONCLUSIONS}
	In this paper, we propose the multiple confidence gates enhancement for joint training of SE and ASR. A speech confidence gates prediction module is designed to replace the former speech enhancement model. And the noisy speech is filtered by each confidence gate respectively. Then the gated results are combined to obtain the input for ASR. With the help of the proposed method, the incompatibility between SE and ASR during joint training is mitigated. And the experimental results show that the proposed method can filter out the non-speech parts without destroying the speech structure, and it's more friendly to ASR during joint training.
	% -------------------------------------------------------------------------
	\bibliographystyle{IEEEbib}
	\bibliography{strings,refs}
	
\end{document}